\documentclass[showpacs,eqsecnum,prd]{revtex4}
\usepackage[dvips]{graphicx}
\usepackage{color}
\usepackage{amsmath}
\newcommand{\Sw}{Schwarzschild }
\newcommand{\RN}{Reissner-Nordstr\"om }
\newcommand{\be}{\begin{equation}}
\newcommand{\ee}{\end{equation}}
\begin{document}

\title{Radiation of the Inner Horizon of the \RN Black Hole}

\author{Ari Peltola}
\email[Electronic address: ]{ari.peltola@phys.jyu.fi} 
\author{Jarmo M\"akel\"a} 
\email[Electronic address: ]{jarmo.makela@phys.jyu.fi}  
\affiliation{Department of Physics, University of Jyv\"askyl\"a, PB 35 (YFL), FIN-40351 Jyv\"askyl\"a, Finland}

\begin{abstract}
Despite of over thirty years of research of the black
hole thermodynamics our understanding of the possible role played by the
inner horizons of \RN and Kerr-Newman black holes in black hole
thermodynamics is still somewhat incomplete: There
are derivations which imply that the temperature of the inner
horizon is negative and it is not quite clear what this
means. Motivated by this problem we perform a detailed analysis of the
radiation emitted by the inner
horizon of the \RN black hole. As a result we find that in a maximally
extended \RN spacetime virtual particle-antiparticle pairs are created
at the inner horizon of the \RN black hole such that real particles
with positive energy and temperature are emitted towards the
singularity from the inner
horizon and, as a consequence, antiparticles with negative energy
are radiated away from the singularity through the inner horizon. We show that
these antiparticles will come out from the white hole horizon in the
maximally extended \RN spacetime, at least when the hole is near
extremality. The energy spectrum of the antiparticles leads to a positive
temperature for the white hole horizon. In other words, our analysis
predicts that in addition to the radition effects of black hole
horizons, also the white hole horizon radiates.
The black hole radiation is caused by the quantum effects
at the outer horizon, whereas the white hole radiation is caused by
the quantum effects at the inner horizon of the \RN black hole.
\end{abstract}

\pacs{04.70.Dy}

\maketitle

\section{Introduction}
Hawking's celebrated paper on black hole radiation \cite{haw} came as
a great surprise to almost everyone working in the field of general
relativity. Till then it was strongly believed that black holes are
totally black, i.e., no matter nor radiation can be emitted by black holes. In
fact, this is what one would expect on the purely
classical grounds. However, when one takes into account the quantum
mechanical effects near the event horizon of a hole, one finds that the
black hole does radiate thermal radiation with a spectrum similar to that of a
black body.

One way to understand the origin of the radiation is to consider
spontaneous particle-antiparticle pair production near the event
horizon of a black hole. Normally, such a pair annihilates itself
very rapidly. It is possible, however, that one of them---particle or
antiparticle---is swallowed by the hole before the annihilation such that the other one is
free to escape away from the hole. This event is illustrated in
Fig. \ref{fig:lakki}.(a) in the case of a \Sw black hole. It can be shown that as a
net effect more antiparticles than particles fall through the horizon
towards the singularity. Therefore an observer outside the hole, that is, at the region I
of the Fig. \ref{fig:lakki}.(a), observes a particle flux which seems to come out from the black
hole.

It is well known that the outer horizon of a Reissner-Nordstr\"om black
hole radiates in a similar way as the event horizon of a \Sw black
hole. However, it is interesting to see what kind of phenomena is
predicted by the virtual pair production mechanism if
one looks at the \emph{inner horizon} of the Reissner-Nordstr\"om black
hole. Consider a
maximally extended Reissner-Nordstr\"om spacetime (see
Fig. \ref{fig:lakki}.(b)). It is easy to see that the causal
relationship between the regions V' and IV' is similar to that between the
regions I and II, respectively. Therefore, as shown in the Fig. \ref{fig:lakki}.(b), a virtual
particle-antiparticle pair which emerges very close to the inner horizon
$r=r_-$ in the region V' can avoid annihilation if either the particle or the antiparticle
falls into the region IV' and the other one remains in V'.
Therefore the pair production mechanism implies that the inner
horizon does radiate and, moreover, that the radiation is directed
\emph{inwards}, towards the singularity. This line of reasoning, however, provides no information about the
radiation itself. Especially, it remains unclear whether the
inner horizon radiates particles or antiparticles.
\begin{figure}
\centering
\input{RNhole3.pstex_t}
\caption{(a) Conformal diagram of the maximally extended \Sw spacetime. In
  this diagram the regions I and III represent spacetime
  surrounding the regions II (black hole) and IV (white
  hole). However, the regions I and III are causally separated. If
  a particle-antiparticle pair is spontaneously created near the event
  horizon of the hole in region I, it is possible that either a particle or an antiparticle is swallowed by the
  hole such that the other one is free to escape to the infinity at
  $\Im^+$. \\ (b) Maximally extended Reissner-Nordstr\"om
  spacetime. Similarly as in the \Sw spacetime, a virtual
  pair created near the inner horizon $r=r_-$
  may avoid annihilation if the particle and the antiparticle are
  separated by the horizon.}
\label{fig:lakki}
\end{figure}

After Hawking's original work there have
been various derivations of the Hawking effect with
different physical assumptions \cite{visser}. Curiously,
very little is known about the radiation of the
inner horizons of the Reissner-Nordstr\"om and the Kerr-Newman black holes. 
This feature can, at least to some extent, be regarded as a
consequence of the fact that inside the inner horizon there are no spacetime regions
analogous to the regions $\Im^+$ or $\Im^-$. After all, Hawking's
original work was based on the analysis of the properties of the Klein-Gordon field at
$\Im^+$ and $\Im^-$. To the best of our knowledge, the only explicit calculation
considering the radiation of the inner horizons has been performed by Wu
and Cai by means of the analytic continuation of the Klein-Gordon
field \cite{wc}. However, as a result of their analysis they found
that the temperature of the inner horizon is \emph{negative} and this
seems to contradict the general attitude towards the black hole
thermodynamics \cite{pad}, as well as the very foundations of
thermodynamics itself. Thus, the true nature of the radiation of the inner horizon
is still somewhat unclear.

The aim of this paper is to perform a detailed analysis of the
radiation of the inner horizon of the \RN black hole.
We would like to point out, however,
that our analysis predicts very little astrophysical
consequences because no mechanism for the formation of \RN
black holes is known. One of the main reasons why the full \RN
spacetime is not considered astrophysically real is the phenomenon
called \emph{mass inflation} near the inner horizon of the \RN black
hole. It is known from the works of Poisson and Israel that when one
considers the spherical collapse of a charged star then, at least in
a somewhat idealized situation, the flux of particles emitted by the
collapsing star and its backscattered counterpart near the inner
horizon of the \RN spacetime provoke an enormous
inflation of the internal mass parameter of the black hole \cite{pi}. Eventually
the mass parameter becomes large enough to form a singularity at the
inner horizon and in effect to freeze the evolution of
spacetime. The inflation of the mass parameter at the inner horizon does not have any
implications in the region outside the black hole since these regions
are causally separated.

In this paper we take the maximally extended \RN spacetime as the
starting point of our analysis. We would like to emphasize that here the
\RN spacetime is considered only as a mathematical solution to the combined
Maxwell-Einstein field equations, and the whole problem
concerning the formation of such a spacetime is completely ignored. One
could ask, of course, why are we interested in the properties of the
full \RN spacetime if it is not considered astrophysically
relevant. The answer to this question lies in the fact that \RN
spacetime provides an explicit example of a spacetime geometry which
contains \emph{two horizons}, of which one is hidden from the outside
observer. The problem we are interested in is the following: Does
only one of the horizons emit Hawking radiation, as it is generally
believed, or do both of the horizons radiate? This is an intriguing
question, and if one is able to show that both of the horizons radiate, then
this result may be seen to support the idea that \emph{all} horizons
of spacetime emit radiation. It is possible that this idea, if true,
may provide useful clues in the search for quantum gravity. 
The main object of interest in our paper is therefore not the \RN
black hole itself, but the general semi-classical properties of
gravity. One may also hope that our results are qualitatively the same for
the more realistic Kerr black hole solution. Indeed, this might be the
case because the causal structures of the \RN and the Kerr spacetimes are very
similar. In that case the radiation of the inner horizon may cause
even certain astrophysical effects since there is not necessarily
mass inflation in the Kerr spacetime.

If the inner horizon of the \RN black hole really radiates, it is
expected that this effect takes place during a very short time only. To see
this, suppose that we begin with the purely classical \RN solution with two
horizons, and apply the results of quantum field theory in the
\RN spacetime. If, in the semiclassical limit, we find that the inner
horizon radiates, then the backscattered part of that radiation is
enough to trigger the mass inflation, and eventually the inner horizon
disappears. In other words, if one is able to show that the inner
horizon of the \RN black hole radiates, then it turns out that semiclassically the full
\RN spacetime, even as a mathematical solution, is unstable. The
radiation process of the inner horizon, however,
should last as long as the inner horizon exists. We shall not discuss
the backscattering effect and its consequences in more detail but we shall confine ourselves
merely to the qualitative aspects discussed above. This approach is
justified because we are interested in the radiation effects of the
inner horizon of the \RN spacetime when the effects of
mass inflation are still neglible.

In brief, the key points of our discussion can be expressed as follows: When the ideas of
Hawking's original work are utilized in Rindler spacetime, one recalls that
the so-called Unruh effect, which is closely related to the Hawking effect, can be obtained by simply comparing the solutions to the
Klein-Gordon equation for massless particles from the points of views
of inertial and uniformly accelerated observers. As a preliminary
we show this in Sec. \ref{sec:pre}. In curved spacetime,
however, the concept of inertial observer is replaced by the concept
of freely falling observer. Inspired by this analogy we proceed to
calculate the effective temperature of black hole horizons
by comparing the solutions to the massless Klein-Gordon equation
from the points of views of an observer in a radial free fall and an observer
at rest with respect to the horizon. First, in Sec.  \ref{sec:outer} we
perform, as an example, an analysis of the radiation of the outer
horizon of a \RN black hole. After reproducing the well-known results,
we proceed, in Sec. \ref{sec:inner}, to calculate the temperature of
the radiation emitted by the inner horizon. In contrast to Wu and Cai,
we find that the effective temperature for particles radiating from
the inner horizon towards the singularity is not negative but
\emph{positive}: The inner horizon emits real particles with positive energy
and temperature. To maintain the local energy balance it is therefore
necessary that the inner horizon emits antiparticles with negative
energy in the direction
away from the singularity. If one looks at the conformal diagram
of a maximally extended \RN spacetime, one may speculate on the possibility that if the
backscattering effects are neglected, the antiparticles emitted
away from the singularity by the inner horizon will go through the intermediate region
between the horizons, and finally they will come out through the white
hole horizon. Motivated by this conjecture, we perform a detailed
analysis of the antiparticle modes, and find that the antiparticles
indeed come out of the white hole---at least when the
black hole is almost extreme. In other words, our analysis predicts a new effect for
maximally extended \RN spacetimes which we shall call ``white hole
radiation''. In the same way as is black hole radiation a consequence
from the quantum effects at the outer horizon of the \RN black hole,
the white hole radiation is a consequence from the quantum effects at
the inner horizon of the hole. The black and the white hole radiations are
separate and simultaneously ongoing processes in spacetimes containing
a \RN black hole, and an observer situated at the exterior region of a
\RN black hole observes the both types of radiation. An existence of
the white hole radiation is the main result of this paper, and it
seems that the same result holds also for Kerr-Newman black holes. We end our
discussion in Sec. \ref{sec:con} with some concluding remarks.

\section{Preliminaries: The Unruh Effect} \label{sec:pre}
\begin{figure}
\centering
\includegraphics{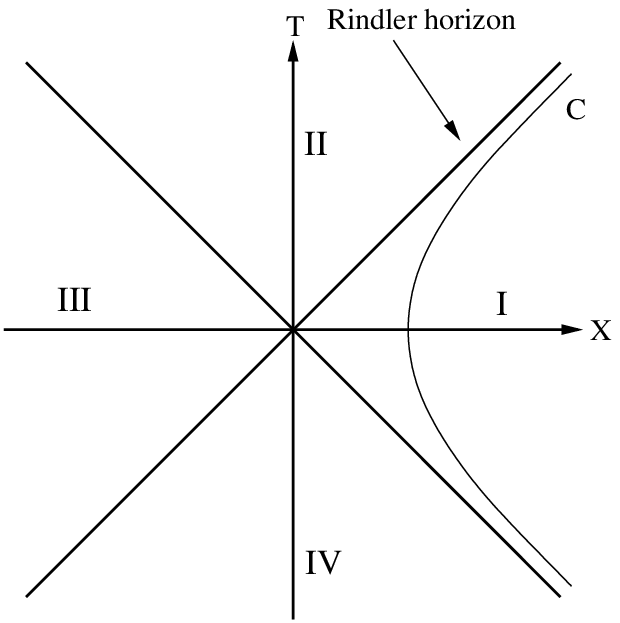}
\caption{Rindler spacetime. The curve $C$ represents the wordline of a
uniformly accelerated observer.}
\label{fig:masa}
\end{figure}

As a starting point of our analysis let us consider the thermal
radiation of the Rindler horizon found by Unruh in 1976 \cite{unruh}.
Rindler horizons are such horizons of spacetime which appear in the
rest frame of a uniformly accelerated observer. In general, the equation of the world 
line of a uniformly accelerated observer in flat two-dimensional
Minkowski spacetime is (unless otherwise stated
we shall always have $c=G=\hbar = k_B = 1$) \cite{bd}:
\begin{equation} X^2-T^2= \frac{1}{a^2},
\end{equation}
where $a$ is the proper acceleration of the observer, and $X$ and $T$, respectively, are the minkowskian space and time coordinates. 
The world line of the observer may be written in the parametrized form:
\begin{subequations} \label{coor}
\be \label{eq:tillin} T(\eta )=\frac{1}{a} \sinh (a\eta ),
\ee
\be \label{eq:tallin} X(\eta )=\frac{1}{a} \cosh (a\eta ).
\ee
\end{subequations}
In this expression, $\eta$ is the proper time of the
observer. If we define the Rindler coordinates $t$ and $x$ such that 
\begin{subequations}
\be x:=\frac{1}{a},
\ee
\be t:=a\eta,
\ee
\end{subequations}
we can write the metric of
two-dimensional Minkowski spacetime as
\be \label{eq:ds} ds^2 =-x^2 dt^2 +dx^2.
\ee
The world line of a uniformly accelerated observer
has been drawn in Fig. \ref{fig:masa}. In that figure we may also see
the Rindler horizon of the accelerated
observer. From the figure we can also see the four regions of Rindler
spacetime, labelled as I, II, III and IV. Since this diagram is very
similar to the Kruskal diagram of \Sw spacetime, one would
expect Rindler spacetime to have physical properties similar to those of
\Sw spacetime. In fact, it is easy to see that the causal
features of the regions II and IV are, respectively, similar to those of a black
and a white hole. 

The simplest way to obtain the Unruh effect is probably the following: Consider the Klein-Gordon equation
of massless particles in the rest frame of an accelerated observer. In general that equation may be
written as
\begin{equation} \label{eq:KGE} g^{\mu \nu}D_\mu D_\nu \phi = 0, 
\end{equation}
where $D_\mu$ denotes covariant derivative, and when spacetime metric is that of Eq. (\ref{eq:ds}),
Eq. (\ref{eq:KGE}) takes the form
\begin{equation} \label{eq:KGE2} \Big(-\frac{1}{x^2} \frac{\partial^2}{\partial t^2}+
\frac{\partial^2}{\partial x^2} +\frac{1}{x}\frac{\partial}{\partial x} \Big) \phi =0.
\end{equation}
If one defines
\begin{equation} x^* := \ln x,
\end{equation}
Eq. (\ref{eq:KGE2}) becomes to:
\begin{equation} \label{eq:KGE3} \Big(-\frac{\partial^2}{\partial t^2}+
\frac{\partial^2}{\partial x^{*2}} \Big) \phi =0.
\end{equation}
Orthonormal solutions to this equation are of the form
\begin{equation} u_\omega = N_\omega e^{-i\omega U},
\end{equation}
where $N_\omega$ is an appropriate normalization constant, and we denote
\begin{equation} \label{eq:Utx} U := t-x^*.
\end{equation}
These solutions represent, from the point of view of an accelerated
observer in the region I, particles
with energy $\omega$ propagating to the positive $X$-direction. In contrast, the corresponding solutions
to the massless Klein-Gordon equation
\begin{equation} \Big( -\frac{\partial^2}{\partial T^2}+
\frac{\partial^2}{\partial X^2} \Big) \phi =0,
\end{equation}
written from the point of view of an inertial observer at rest with respect to the Minkoski
coordinates $T$ and $X$, are of the form:
\begin{equation} u_\omega ' = N_\omega e^{-i\omega \widetilde{u}},
\end{equation}
where
\begin{equation} \widetilde{u} := T-X.
\end{equation}
Again, these solutions represent particles with energy $\omega$
propagating to the positive $X$-direction. 

\begin{figure}
\centering
\includegraphics{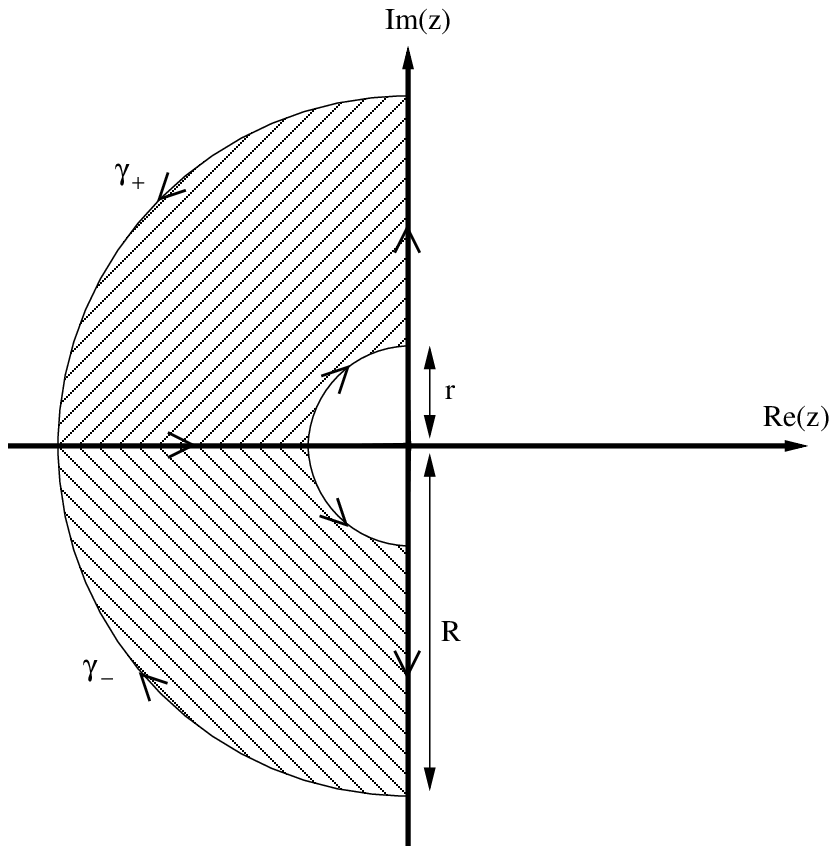}
\caption{Integration contours in the complex plane. In this figure
  $\gamma_+$ and $\gamma_-$ are closed contours circulating
  the shaded regions in the upper and the lower half of the complex
  plane, respectively. When the integral in Eq. (\ref{eq:picard})
  is calculated along the contour $\gamma_+$, one easily sees that, in
  the limit where $R \rightarrow \infty$ and $r \rightarrow 0$, the
  integrals along the arcs of the circles vanish. The analyticity
  of the functions under consideration in the shaded regions implies
  that the contour integral around $\gamma_+$ vanishes,
  and therefore the integral from negative infinity to zero along the
  real axis may be transformed into an integral from positive infinity
  to zero along the imaginary axis. Similar result holds for the integral in
  Eq. (\ref{eq:data}) along the path $\gamma_-$, except that now the
  integral from negative infinity to zero along the real axis may be
  transformed to an integral from negative infinity to zero along the
  imaginary axis. The integrals along the
  imaginary axis lead directly to Eq. (\ref{eq:iiips}).}
\label{fig:complex}
\end{figure}

It is easy to see from
Eqs. (\ref{coor}) and (\ref{eq:Utx}) that
\begin{equation} U = - \ln (-\widetilde{u}),
\end{equation}
and therefore the Bogolubov transformation
\begin{equation} u_\omega = \sum_{\omega '} \Big( A_{\omega \omega '}' u_{\omega '}'
+B_{\omega \omega '}'u_{\omega'}'^* \Big)
\end{equation}
between the orthonormal solutions $u_\omega$ and $u_\omega '$ may be written as
\begin{equation}  e^{i\omega \ln (-\widetilde{u})}= \sum_{\omega '} \Big( A_{\omega \omega '}'
e^{-i\omega' \widetilde{u}} + B_{\omega \omega '}' e^{i\omega '\widetilde{u}} \Big),
\end{equation}
where the Bogolubov coefficients $A_{\omega \omega '}'$ and $B_{\omega \omega '}'$ are expressible 
as Fourier integrals:
\begin{subequations} \label{molli}
\begin{equation} \label{eq:picard} A_{\omega \omega '}' = \frac{1}{2\pi} \int\limits_{-\infty}^0 d\widetilde{u}\,  
e^{i\omega \ln (-\widetilde{u})} e^{i\omega '\widetilde{u}},
\end{equation}
\begin{equation} \label{eq:data} B_{\omega \omega '}' = \frac{1}{2\pi} \int\limits_{-\infty}^0 d\widetilde{u}\, 
e^{i\omega \ln (-\widetilde{u})} e^{-i\omega '\widetilde{u}}.
\end{equation}
\end{subequations}
The integration is performed from the negative infinity to zero because we are considering
particles in the region I, and in that region $\widetilde{u}<0$. It is straightforward to show, by performing
the integration in the complex plane, that
\begin{equation} \label{eq:iiips} \big| A_{\omega \omega '}' \big| = e^{\pi \omega} \big| B_{\omega \omega '}' \big|
\end{equation}
(see Fig. \ref{fig:complex}). Because of the well-known relation between the Bogolubov coefficients,
\begin{equation} \label{eq:pernaa}\sum_{\omega '} \Big( \big| A_{\omega \omega '}' \big|^2 -
\big| B_{\omega \omega '}' \big|^2 \Big) = 1,
\end{equation}
one finds that when the field is in vacuum from the point of view of
an inertial observer, the number of particles with energy $\omega$ is,
from the point of view of an accelerated observer,
\begin{equation} \label{eq:n} n_\omega = \sum_{\omega '} \big| B_{\omega \omega '}' \big|^2 =
\frac{1}{e^{2 \pi \omega} - 1}.
\end{equation}
This is the Planck spectrum at the temperature $T_0 = \frac{1}{2\pi}$,
which is related to the temperature experienced by an observer
situated at a given point in space by the Tolman relation \cite{landau}:
\be \label{eq:tolman}T = (g_{00})^{-\frac{1}{2}}\, T_0.
\ee
Hence it follows that a uniformly accelerated observer observes particles
coming out from the Rindler horizon with the black-body spectrum
corresponding to the characteristic temperature
\begin{equation} T_U :=
\frac{1}{2\pi x} = \frac{a}{2\pi},
\end{equation}
even when, from the point of view of an inertial observer, the field
is in vacuum. This result is known as the \emph{Unruh effect}, and it
is one of the most remarkable outcomes of quantum field theory.

\section{Reconsideration of the Hawking Effect for the Outer Horizon
  of a \RN Black Hole} \label{sec:outer}
The ideas inspired by the properties of
Rindler spacetime can be easily utilized when one investigates the
thermal properties of \RN horizons in the following manner: At first
one constructs a certain \emph{geodesic system of coordinates} for the neighborhood of
the horizon under scrutiny. The geodesic coordinates are constructed
such that an observer in a \emph{radial} free fall remains at rest with respect to
those coordinates. Such an observer does not observe the horizon,
and therefore one expects that no radiation
effects should be experienced by him. Because of that one may view the particle vacuum of
the freely falling observer as the vacuum that would exist in spacetime if there were
no horizon at all. In this sense the observer in a radial free fall is
analogous to the inertial observer in flat spacetime, and
we take this similarity as a starting point of our analysis.

To calculate the particle flux emitted by the horizon one compares the
solutions to the massless Klein-Gordon equation
very close to the horizon in two different coordinate systems. One of
these systems is the geodesic system of coordinates and the
other one
is a coordinate system at rest with respect to the horizon. The
analysis of the Klein-Gordon modes must be performed infinitesimally
close to the horizon for the very reason that only in that case  one
is able to solve the Klein-Gordon equation analytically (excluding, of
course, the solutions at the asymptotic infinities). Then
one can obtain the Bogolubov transformations between these solutions
and infer the effective temperature of the radiation flux from the
point of view of an observer at rest with respect to the horizon.

To see what all this really means consider, as an example, the outer horizon
of a \RN black hole. The \RN metric can be written as
\begin{equation} ds^2= -\Bigg(1-\frac{2M}{r} +\frac{Q^2}{r^2} \Bigg)
  dt^2 + \Bigg( 1- \frac{2M}{r} +\frac{Q^2}{r^2} \Bigg)^{-1} dr^2 + r^2(d\theta^2 + \sin^2 \theta\, d\varphi^2), 
\end{equation}
where $M$ is the mass and $Q$ is the electric charge of the
hole. In addition to the physical singularity at $r=0$, this metric
has two coordinate singularities when 
\be r= r_\pm := M \pm \sqrt{M^2-Q^2}.
\ee
The two-surfaces where $r=r_+$ and $r=r_-$ are called, respectively,
the outer and the inner horizons of the \RN black hole.
It is well known that when $r > r_+$, and the backscattering effects
are neglected, an observer at rest with respect to the coordinates $r$, $\theta$, and $\varphi$
observes thermal radiation emitted by the hole with a characteristic temperature
\begin{equation} \label{eq:dilli} T_+ := \frac{\kappa_+}{2\pi \sqrt{1-
      \frac{2M}{r} +\frac{Q^2}{r^2}}} = \frac{\sqrt{M^2-Q^2}}{2\pi
      \big( M^2 + \sqrt{M^2-Q^2}\big)^2 \sqrt{1- \frac{2M}{r} +\frac{Q^2}{r^2}} },
\end{equation}
where 
\be \kappa_+ := \frac{r_+ -r_-}{2 r_+^2}
\ee 
is the surface gravity of the outer horizon. The
factor $(1- \frac{2M}{r} +\frac{Q^2}{r^2})^{-1/2}$ is due to the
``redshift''  of the radiation. At asymptotic infinity the redshift
factor is equal to one, but at the horizon $r=r_+$ it becomes
infinitely large. In other words, an observer at rest very close
to the horizon may measure an infinite temperature for the
black hole radiation. 

We shall now show how Eq. (\ref{eq:dilli}) may be obtained by means of
the method we explained at the beginning of this section. As the first
step we separate the Klein-Gordon field $\phi$ of massless particles
such that
\be \phi (t,r,\theta ,\varphi) = \frac{1}{r}\, f(t,r)\, Y_{lm}(\theta ,\varphi),
\ee
where $Y_{lm}$ is the spherical harmonic function satisfying the
differential equation
\be \Bigg[ \frac{1}{\sin \theta} \frac{\partial}{\partial \theta} \Bigg( \sin
  \theta \frac{\partial}{\partial \theta} \Bigg) +\frac{1}{\sin^2 \theta}
  \frac{\partial^2}{\partial \varphi^2} \Bigg]Y_{lm} = -l(l+1)Y_{lm},
\ee
where, as usual, the allowed values of $l$ are $0,1,2,\cdots ,$ and
those of $m$ are $0, \pm 1,\pm 2, \cdots, \pm l$.
In that case the massless Klein-Gordon equation, 
when written in terms of the
coordinates $t$, $r$, $\theta$, and $\varphi$, implies that
\be \label{eq:reima} \Bigg[ -\frac{\partial^2}{\partial t^2} +\frac{\partial^2}{\partial
r_*^2}-V(r) \Bigg] f = 0,
\ee
where we have defined the ``tortoise coordinate'' $r_*$ such that
\be \label{eq:kilpi} r_* := \int \frac{dr}{1-\frac{2M}{r} +\frac{Q^2}{r^2}} = r -
\frac{r_-^2}{r_+ -r_-} \ln | r-r_- | + \frac{r_+^2}{r_+ -r_-} \ln | r-r_+ |,
\ee
and the ``potential term''
\be \label{eq:leksa} V(r) := \Bigg(1-\frac{2M}{r} +\frac{Q^2}{r^2} \Bigg) \Bigg[
\frac{l(l+1)}{r^2} +\Bigg( \frac{2M}{r^3} - \frac{2Q^2}{r^4} \Bigg) \Bigg]. 
\ee
Very close to the horizon, where
\be \Delta := \sqrt{1-\frac{2M}{r} +\frac{Q^2}{r^2}} 
\ee
is infinitesimally small, the potential $V(r)$
vanishes, and the solutions to Eq. (\ref{eq:reima})
corresponding to the particles with energy $\omega$ moving towards
the horizon are of the form
\be f(t,r) \sim e^{-i\omega V},
\ee 
and for the solutions coming outwards from the horizon we have
\be f(t,r) \sim e^{-i\omega U},
\ee 
where the coordinates $V$ and $U$ are the advanced and
the retarded coordinates defined as: 
\begin{subequations}
\be V = t+r_*,
\ee 
\be \label{eq:U} U=t-r_*.
\ee 
\end{subequations}
Therefore, from the point of view of the observer at rest very close
to the horizon, the ingoing and the outcoming
solutions to the massless Klein-Gordon equation are, respectively,
\begin{subequations} \label{palikka}
\be \label{eq:sisaan} \phi_{\text{in}} \approx N_{\omega lm} Y_{lm}\, \frac{1}{r}\,  e^{-i\omega V},
\ee
\be \label{eq:ulos} \phi_{\text{out}} \approx N_{\omega lm} Y_{lm}\, \frac{1}{r}\,  e^{-i\omega U},
\ee
\end{subequations}
where $N_{\omega lm}$ is an appropriate normalization constant.

\begin{figure}
\centering
\input{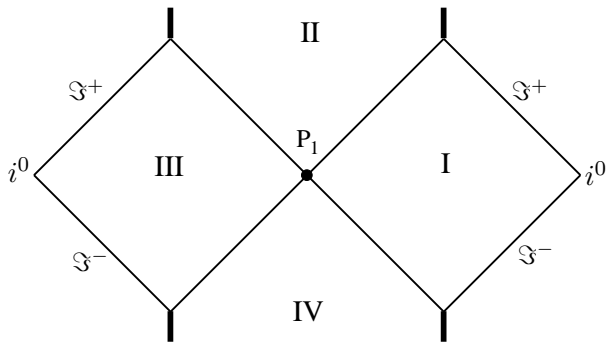}
\caption{Part of \RN spacetime. The bifurcation point $P_1$ is situated at the
  intersection of the lines that separate the regions I, II, III and IV.}
\label{fig:lasse}
\end{figure}
Consider now an observer in a radial free fall in the region I of the \RN spacetime,
infinitesimally close to the point $P_1$
in Fig. \ref{fig:lasse}. 
In other words, the observer is in a radial free fall just outside
the outer horizon. At first let us introduce in the \RN 
spacetime the coordinates $(u,v)$ which are similar to the Kruskal coordinates in \Sw
spacetime. In general, these ``Kruskal-type coordinates'' may be defined in
the different regions of the \RN spacetime such that
\begin{subequations} \label{ozzy}
\begin{eqnarray} \label{eq:ritmo} \left\{ \begin{array}{l}
u = \frac{1}{2} \big( e^{\alpha V} + e^{-\alpha U} \big), \\ 
v =\frac{1}{2} \big( e^{\alpha V} - e^{-\alpha U} \big),  \end{array}
\right. & \qquad \qquad & (\textrm{Region I, I',} \cdots ) \\
\label{eq:ritmo55} \left\{ \begin{array}{l}
u = \frac{1}{2} \big( e^{\alpha V} - e^{-\alpha U} \big), \\ 
v =\frac{1}{2} \big( e^{\alpha V} + e^{-\alpha U} \big),  \end{array}
\right. & \qquad \qquad & (\textrm{Region II, II',} \cdots ) \\
\label{eq:ritmo66} \left\{ \begin{array}{l}
u = -\frac{1}{2} \big( e^{\alpha V} + e^{-\alpha U} \big), \\ 
v = -\frac{1}{2} \big( e^{\alpha V} - e^{-\alpha U} \big),  \end{array}
\right. & \qquad \qquad & (\textrm{Region III, III',} \cdots ) \\
\label{eq:ritmo2} \left\{ \begin{array}{l}
u = -\frac{1}{2} \big( e^{\alpha V} - e^{-\alpha U} \big), \\ 
v = -\frac{1}{2} \big( e^{\alpha V} + e^{-\alpha U} \big),  \end{array}
\right. & \qquad \qquad & (\textrm{Region IV, IV',} \cdots )
\end{eqnarray}
\end{subequations}
where $\alpha$ is an appropriate constant. When we study the physical
properties of 
the outer horizon, the constant $\alpha$ must be
chosen such that the metric on the two-surface $r=r_+$ is regular. The
most natural choice is 
\be \alpha = \kappa_+,
\ee
and this choice leads to the non-singular metric 
\be ds^2 = \frac{1}{\kappa_+^2 r^2} e^{-2\kappa_+ r} (r-r_-
)^{\frac{r_-^2}{r_+^2}+1} (-dv^2 +du^2 ) + r^2(d\theta^2 + \sin^2
\theta\, d\varphi^2).
\ee 

It is now easy to construct a geodesic coordinate
system for an infinitesimal neighborhood $\mathcal{U}(P_1)$ of the
point $P_1$.
By infinitesimal geodesic coordinate system we mean coordinates
$X^I$ ($I=0,1,2,3$) in $\mathcal{U}(P_1)$ such that, at the point $P_1$, the metric
takes the form of that of flat spacetime, i.e.,
\be \label{eq:tuire} ds^2 = \eta_{IJ}\, dX^I dX^J,
\ee
where $\eta_{IJ} = \text{diag} (-1,1,1,1)$ is the flat Minkowski metric 
and the derivatives of the metric vanish. Let us define the coordinates
\begin{subequations} \label{roisto}
\be \label{eq:riippi} X^0 := l_+ v,
\ee
\be \label{eq:kauppila} X^1 := l_+ u,
\ee
\end{subequations}
where 
\be \label{eq:worf} l_+ := \frac{1}{\kappa_+ r_+} e^{-\kappa_+ r_+} (r_+ -
r_-)^{\frac{1}{2} \big( \frac{r_-^2}{r_+^2} +1 \big) }.
\ee
By using these definitions one finds that at the point $P_1$ the metric can be written as
\be \label{eq:rekku} ds^2 =-\big( dX^0 \big)^2 +\big( dX^1 \big)^2 + r^2(d\theta^2 + \sin^2 \theta\, d\varphi^2),
\ee
and the derivatives of the metric with respect to $X^0$ and $X^1$
vanish (for details, see Appendix). Therefore, the geodesic
coordinates of the freely falling observer can be chosen to be $X^0$ and $X^1$.
(Note that even though the above metric is not stricly of the form of
Eq. (\ref{eq:tuire}), for the observer in a radial free fall these
coordinates provide a geodesic coordinate system since in that case
$\theta$ and $\varphi$ are constants.)

If the massless Klein-Gordon field is now separated such that 
\be \label{eq:moppe} \phi (X^0, X^1,\theta ,\varphi) = \frac{1}{r}\, \widetilde{f}(X^0,X^1)\, Y_{lm}(\theta ,\varphi),
\ee
the Klein-Gordon equation, when written in terms of the coordinates
$X^0$, $X^1$, $\theta$, and $\varphi$, implies:
\be \label{eq:reima2} \Bigg[ -\frac{\partial^2}{\partial \big( X^0 \big)^2} +\frac{\partial^2}{\partial
\big( X^1\big)^2} + \frac{1}{r} \Bigg( \frac{\partial^2 r}{\partial
\big( X^0 \big)^2} -\frac{\partial^2 r}{\partial \big( X^1 \big)^2}
\Bigg)- \frac{l(l+1)}{r^2} F_+(r) \Bigg] \widetilde{f}(X^0, X^1) = 0,
\ee
where we have denoted:
\be F_+ (r) := \frac{1}{\kappa_+^2 l_+^2}\frac{1}{r^2} e^{-2\kappa_+
  r} (r-r_-)^{\frac{r_-^2}{r_+^2}+1}. 
\ee
It follows from Eqs. (\ref{roisto}),
(\ref{eq:viikate}) and (\ref{eq:viikate2}) that
\begin{subequations}
\be \frac{\partial r}{\partial X^0} = -\kappa_+ F_+ (r) X^0, 
\ee
\be \frac{\partial r}{\partial X^1} = \kappa_+ F_+ (r) X^0, 
\ee
\end{subequations}
and therefore:
\be \frac{\partial^2 r}{\partial
\big( X^0 \big)^2} -\frac{\partial^2 r}{\partial \big( X^1 \big)^2} =
\kappa_+ F_+ (r) \Big\{ \kappa_+ F_+' (r) \big[ \big( X^0\big)^2 -\big(
  X^1\big)^2\big] -2 \Big\} ,
\ee
where the prime means derivative with respect to $r$.
Using Eqs. (\ref{eq:previi})--(\ref{eq:viikate2}) and (\ref{eq:kilpi})
one finds:
\be F_+' (r) \big[ \big( X^0\big)^2 -\big(
  X^1\big)^2\big]  = \frac{2}{\kappa_+} + \frac{1}{\kappa_+^2} \Bigg(
  \frac{2M}{r^2} - \frac{2Q^2}{r^3}\Bigg),
\ee
and therefore Eq. (\ref{eq:reima2}) takes the form:
\be \label{eq:reima5} \Bigg[ -\frac{\partial^2}{\partial \big( X^0 \big)^2} +\frac{\partial^2}{\partial
\big( X^1\big)^2} - \widetilde{V}(r) \Bigg] \widetilde{f}(X^0, X^1) = 0,
\ee
where the ``potential term'' is
\be \widetilde{V}(r) := \Bigg[ -\frac{2M}{r^3} +\frac{2Q^2}{r^4}+
\frac{l(l+1)}{r^2}\Bigg] F_+ (r).
\ee
The function $F_+ (r)$ has the property
\be F_+ (r_+) = 1,
\ee
and therefore Eq. (\ref{eq:reima5}) takes, at the outer horizon of the
\RN black hole, the form:
\be \label{eq:reima6} \Bigg[ -\frac{\partial^2}{\partial \big( X^0 \big)^2} +\frac{\partial^2}{\partial
\big( X^1\big)^2} + \frac{2M}{r^3_+} -\frac{2Q^2}{r^4_+}-
\frac{l(l+1)}{r^2_+}\Bigg] \widetilde{f} (X^0, X^1)= 0.
\ee
So we see that, in contrast to Eq. (\ref{eq:reima}), the ``potential
term'' does not vanish at the horizon. For a macroscopic hole, however,
the ``potential term'' may be neglected: For \RN black holes $r_+ \geq
M$ and $0 \leq |Q| \leq M$, and so it follows that
\be \Bigg| \frac{2M}{r^3_+} -\frac{2Q^2}{r^4_+} \Bigg| \leq
\frac{2}{M^2},
\ee
which means that when, in Planck units, $M \gg 1$, the terms involving
$M$ and $Q$ will vanish. Moreover, if the orbital angular momentum $l$
of the Klein-Gordon particle is sufficiently small, we may neglect the
term $l(l+1)/r_+^2$. In other words, we may write
Eq. (\ref{eq:reima6}), in effect, as:
\be \label{eq:reima7} \Bigg[ -\frac{\partial^2}{\partial \big( X^0
  \big)^2} +\frac{\partial^2}{\partial \big( X^1 \big)^2} \Bigg]
\widetilde{f}(X^0, X^1) = 0.
\ee

For very small $l$ the ingoing and the outcoming
solutions to the massless Klein-Gordon equation very close to the
horizon $r=r_+$ are:
\begin{subequations}
\be \label{eq:sisaan2} \phi_{\text{in}}' \approx N_{\omega lm} Y_{lm}\, \frac{1}{r}\,  e^{-i\omega \widetilde{v}},
\ee
\be \label{eq:ulos2} \phi_{\text{out}}' \approx N_{\omega lm} Y_{lm}\, \frac{1}{r}\,  e^{-i\omega \widetilde{u}},
\ee
\end{subequations}
where
\begin{subequations} \label{juusot}
\be \label{eq:juuso} \widetilde{v} = X^0 +X^1,
\ee
\be \label{eq:wille} \widetilde{u} = X^0 -X^1,
\ee
\end{subequations}
and $N_{\omega lm}$ is the normalization constant corresponding to the
fixed values of $l$, $m$, and $\omega$.
These solutions represent, from the point of view of the freely
falling observer, particles with energy $\omega$ moving towards and
out of the horizon, respectively. From Eqs. (\ref{eq:U}), (\ref{roisto}),
and (\ref{eq:wille}) one easily finds that in
the region I 
\be \label{eq:makkara}
U = - \kappa_+^{-1} \ln (-\widetilde{u}) + \kappa_+^{-1} \ln l_+.
\ee
Thus, in the spherical symmetric case, the
Bogolubov transformation between the outcoming modes in Eqs. (\ref{eq:ulos}) and
(\ref{eq:ulos2}) can be written in the form
\begin{equation}  e^{i\omega \kappa_+^{-1}\ln (-\widetilde{u})}\, e^{-i\omega \kappa_+^{-1} \ln l_+} = \sum_{\omega '} \Big( A_{\omega \omega '}'
e^{-i\omega' \widetilde{u}} + B_{\omega \omega '}' e^{i\omega '\widetilde{u}} \Big),
\end{equation}
and, moreover, we can express the Bogolubov coefficients $A_{\omega \omega '}'$ and $B_{\omega \omega '}'$
as Fourier integrals such that
\begin{subequations} \label{duuri}
\begin{equation} \label{eq:kala} A_{\omega \omega '}' = \frac{1}{2\pi}
  e^{-i\omega \kappa_+^{-1} \ln l_+}  \int\limits_{-\infty}^0 d\widetilde{u}\,  
e^{i\omega \kappa_+^{-1}\ln (-\widetilde{u})} e^{i\omega '\widetilde{u}},
\end{equation}
\begin{equation} \label{eq:kukko} B_{\omega \omega '}' = \frac{1}{2\pi} e^{-i\omega \kappa_+^{-1} \ln l_+}\int\limits_{-\infty}^0 d\widetilde{u}\, 
e^{i\omega \kappa_+^{-1}\ln (-\widetilde{u})} e^{-i\omega '\widetilde{u}}.
\end{equation}
\end{subequations}
As in the previous section, the integration is performed from the negative infinity to zero because we are considering
particles in the region I, and in that region
$\widetilde{u}<0$. 

The integrals in Eqs. (\ref{duuri}) are similar to those
found in Eqs. (\ref{molli}), and the integration in the complex plane gives
\begin{equation} \big| A_{\omega \omega '}' \big| = e^{\pi \kappa_+^{-1}\omega} \big| B_{\omega \omega '}' \big|.
\end{equation}
Therefore, by using Eq. (\ref{eq:pernaa}), we find that when the field
is in vacuum from the point
of view of a freely falling observer, the
number of the particles with energy $\omega$ observed by
an observer at rest very close to the horizon is 
\begin{equation} n_\omega = \sum_{\omega '} \big| B_{\omega \omega '}' \big|^2 =
\frac{1}{e^{2 \pi \kappa_+^{-1}\omega} - 1}.
\end{equation}
This is the Planck spectrum at the temperature 
\be T = \frac{\kappa_+}{2\pi},
\ee
which represents the temperature of the outer horizon experienced by an observer at rest with respect
to the horizon when the redshift effects of the radiation are
ignored. The redshift factor can be recovered
by the Tolman relation (\ref{eq:tolman}), and as a result we find that,
from the point of view of an observer at rest very close
to the outer horizon, the \RN 
black hole emits radiation with a characteristic temperature
\begin{equation} \label{eq:dilli2} T_+ = (g_{00})^{-\frac{1}{2}} \frac{\kappa_+}{2\pi} = 
\frac{\kappa_+}{2\pi \Delta} = \frac{\sqrt{M^2-Q^2}}{2\pi r_+^2
  \sqrt{1-\frac{2M}{r} +\frac{Q^2}{r^2}}},
\end{equation}
which is Eq. (\ref{eq:dilli}). In other words, we have shown that our
method, which relies on the comparison of the solutions to the
Klein-Gordon equation from the points of views of two observers close
to the horizons, reproduces the familiar result which is usually
obtained by means of the comparison of the solutions to the
Klein-Gordon equation at $\Im^+$ and $\Im^-$. Encouraged by this
welcome outcome of our analysis we now proceed to apply our method for
an analysis of the properties of the inner horizon of the \RN black hole.

\section{Hawking Effect for the Inner Horizon of the \RN Black Hole} \label{sec:inner}
An analysis of the radiation emitted by the inner horizon of a \RN black hole can
now be performed in a very similar way as that of the outer
horizon. In essence, the ingoing and the outcoming solutions to the
Klein-Gordon equation, when written in terms of the coordinates $t$,
$r$, $\theta$, and $\varphi$, can be obtained directly from
Eqs. (\ref{palikka}). However, since the
radiation is now directed towards the singularity $r=0$, the observer at rest with respect
to the inner horizon must be situated \emph{inside} the two-sphere
$r=r_-$. Therefore the roles of the ingoing and the outcoming modes
interchange. More precisely, the solutions representing particles with
energy $\omega$ are,
from the point of view of the observer at rest very close to the inner
horizon,
\begin{subequations}
\be \label{eq:sisaan3} \phi_{\text{in}} \approx N_{\omega lm} Y_{lm}\, \frac{1}{r}\,  e^{-i\omega U},
\ee
\be \label{eq:ulos3} \phi_{\text{out}} \approx N_{\omega lm} Y_{lm}\, \frac{1}{r}\,  e^{-i\omega V}.
\ee
\end{subequations}
The solution $\phi_{\text{in}}$ represents a particle which moves
towards the horizon, and therefore away from the singularity. The
solution $\phi_{\text{out}}$, in turn, represents a particle which
moves out of the horizon, and therefore towards the singularity.

\begin{figure}
\centering
\input{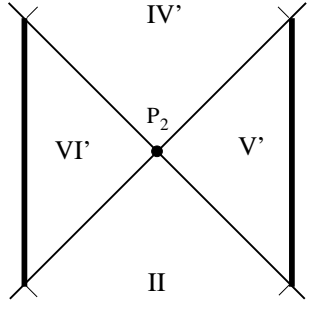}
\caption{Part of \RN spacetime. The point $P_2$ is situated at the
  intersection of the lines that separate the regions V', IV', VI' and II.}
\label{fig:lasse2}
\end{figure}

As it comes to the freely falling observer near the inner horizon, we cannot use the same
geodesic coordinate system as we did in the previous section. This is a
consequence of the fact that the Kruskal-type coordinates $u$ and $v$
of Eqs. (\ref{ozzy})
with the choice $\alpha=\kappa_+$ lead to the metric which is not
regular at $r=r_-$. A remedy to this problem can be obtained by
choosing 
\be \alpha = \kappa_- := -\frac{r_+ -r_-}{2 r_-^2}
\ee
and defining
a new geodesic system of coordinates based on this choice. 
Consider now an observer in a radial free fall in the region VI'
of \RN spacetime infinitesimally close to the point $P_2$ 
(see Fig. \ref{fig:lasse2}). When
written in terms of the coordinates $u$ and $v$, the spacetime metric
takes the form:
\be ds^2 = \frac{1}{\kappa_-^2 r^2} e^{-2\kappa_- r} (r_+ - r
)^{\frac{r_+^2}{r_-^2}+1} (-dv^2 +du^2 ) + r^2(d\theta^2 + \sin^2
\theta\, d\varphi^2).
\ee
In this expression the coordinates $u$ and $v$ have been defined in
such a way that in the regions V', IV', VI' and II of
Fig. \ref{fig:lakki}.(b), respectively, the coordinates $u$ and $v$
are given in terms of $U$ and $V$ by Eqs. (\ref{eq:ritmo}), (\ref{eq:ritmo55}),
(\ref{eq:ritmo66}), and (\ref{eq:ritmo2}). One may also easily check that
in the region VI' the coordinates $u$ and $v$ are increasing functions
of $r$ and $t$, respectively. More precisely, when $u$ is taken to be
a constant, the coordinate $v$ increases as a function of $t$, whereas
the coordinate $u$ increases as a function of $r$, when $v$ is constant. 

Similarly as in the case of the
outer horizon, we define a geodesic system of coordinates for an
infinitesimal neighborhood of the point $P_2$ such that 
\begin{subequations} \label{verneri}
\be \label{eq:riippi2} X^0 := l_- v,
\ee
\be \label{eq:kauppila2} X^1 := l_- u,
\ee
\end{subequations}
where 
\be \label{eq:worf2} l_- := \frac{1}{|\kappa_-| r_-} e^{-\kappa_- r_-} (r_+ -
r_-)^{\frac{1}{2}\big( \frac{r_+^2}{r_-^2} +1\big) }.
\ee
When the remaining coordinates are chosen to be the spherical
coordinates $\theta$ and $\varphi$, the metric is given by
Eq. (\ref{eq:rekku}), and the derivatives of the metric vanish when
$r=r_-$ (see
Appendix). Therefore the coordinates $X^0$ and $X^1$ provide a geodesic
system of coordinates. Furthermore, when the massless Klein-Gordon field
is separated as in Eq. (\ref{eq:moppe}), one finds that the massless
Klein-Gordon equation implies:
\be \label{eq:raimo} \Bigg[ -\frac{\partial^2}{\partial \big( X^0 \big)^2} +\frac{\partial^2}{\partial
\big( X^1\big)^2} + \Bigg( \frac{2M}{r^3} -\frac{2Q^2}{r^4}-
\frac{l(l+1)}{r^2}\Bigg) F_- (r) \Bigg] \widetilde{f} (X^0, X^1)= 0,
\ee
where we have defined:
\be F_- (r) :=  \frac{1}{\kappa_-^2 l_-^2}\frac{1}{r^2} e^{-2\kappa_-
  r} (r_+ -r)^{\frac{r_+^2}{r_-^2}+1}. 
\ee
Again, one finds that 
\be F_- (r_-) =1,
\ee
and therefore Eq. (\ref{eq:raimo}) takes, at the point $P_2$, the
form
\be \label{eq:raimo2} \Bigg[ -\frac{\partial^2}{\partial \big( X^0 \big)^2} +\frac{\partial^2}{\partial
\big( X^1\big)^2} + \frac{2M}{r^3_-} -\frac{2Q^2}{r^4_-} -
\frac{l(l+1)}{r^2_-} \Bigg] \widetilde{f} (X^0, X^1)= 0.
\ee

The question about whether the terms involving $r_-$ are negligibly
small or not, is a very delicate one, and when the absolute value of
the electric charge $Q$ is very small, those terms will certainly
\emph{not} vanish. We may, however, consider a special case where
$|Q|$ is ``reasonably big''. More precisely, we shall assume that
there is a fixed positive number $\gamma \leq 1$ such that between
$|Q|$ and $M$ there is, in Planck units, the relationship:
\be |Q| = \gamma M.
\ee
In that case
\be r_- = (1-\sqrt{1-\gamma^2})M,
\ee
and because
\be \sqrt{1-\gamma^2} = 1 -\frac{1}{2}\gamma^2 - \frac{1}{8}\gamma^4
-\frac{1}{16} \gamma^6 -\cdots < 1-\frac{1}{2}\gamma^2,
\ee
we find that
\be \Bigg|  \frac{2M}{r^3_-} -\frac{2Q^2}{r^4_-} \Bigg| \leq
\frac{2M}{r^3_-} + \frac{2Q^2}{r^4_-} < \frac{48}{\gamma^6} \frac{1}{M^2}.
\ee
Hence it follows that if $\gamma$ is ``reasonably big'', and $M\gg 1$
in Planck units, the terms involving $M$ and $Q$ are negligible. The
same line of reasoning implies that for ``sufficiently small'' $l$ the
term $l(l+1)/r_-^2$ may be neglected, and Eq. (\ref{eq:raimo2}) may be
written, in effect, in the form:
\be \Bigg[ -\frac{\partial^2}{\partial \big( X^0 \big)^2} +\frac{\partial^2}{\partial
\big( X^1\big)^2}  \Bigg] \widetilde{f} (X^0, X^1)= 0.
\ee
One easily sees that in the region VI' the solutions 
corresponding to the particles with energy $\omega$ going in and out
of the horizon are
\begin{subequations}
\be \label{eq:sisaan4} \phi_{\text{in}}' \approx N_{\omega lm} Y_{lm}\, \frac{1}{r}\,  e^{-i\omega \widetilde{u}},
\ee
\be \label{eq:ulos4} \phi_{\text{out}}' \approx N_{\omega lm} Y_{lm}\, \frac{1}{r}\,  e^{-i\omega \widetilde{v}},
\ee
\end{subequations}
where $\widetilde{v}$ and $\widetilde{u}$ are defined as in
Eqs. (\ref{juusot}).

It is now possible to write the Bogolubov transformation between the
outcoming solutions (\ref{eq:ulos3}) and (\ref{eq:ulos4}) in the
spherical symmetric case. One easily
finds that in the region VI',
\be \label{eq:kissa} V = \kappa_-^{-1} \ln (-\widetilde{v}) -
\kappa_-^{-1} \ln l_- .
\ee
Moreover, the Bogolubov transformation takes the form
\begin{equation}  e^{-i\omega \kappa_-^{-1}\ln (-\widetilde{v})}\, e^{i\omega \kappa_-^{-1} \ln l_-} = \sum_{\omega '} \Big( A_{\omega \omega '}'
e^{-i\omega' \widetilde{v}} + B_{\omega \omega '}' e^{i\omega '\widetilde{v}} \Big),
\end{equation}
and the Bogolubov coefficients
can be expressed as:
\begin{subequations}
\begin{equation} A_{\omega \omega '}' = \frac{1}{2\pi} e^{i\omega
    \kappa_-^{-1} \ln l_-} \int\limits_{-\infty}^0 d\widetilde{v}\,  
e^{-i\omega \kappa_-^{-1}\ln (-\widetilde{v})} e^{i\omega '\widetilde{v}},
\end{equation}
\begin{equation} B_{\omega \omega '}' = \frac{1}{2\pi} e^{i\omega \kappa_-^{-1} \ln l_-}\int\limits_{-\infty}^0 d\widetilde{v}\, 
e^{-i\omega \kappa_-^{-1}\ln (-\widetilde{v})} e^{-i\omega '\widetilde{v}}.
\end{equation}
\end{subequations}
As before, these integrals yield the result
\begin{equation} \big| A_{\omega \omega '}' \big| = e^{-\pi \kappa_-^{-1}\omega} \big| B_{\omega \omega '}' \big|.
\end{equation} 
Therefore, by using Eq. (\ref{eq:pernaa}), we see that the
number of particles with energy $\omega$ is, from the point of view of
the observer at rest very close to the horizon $r=r_-$,
when, from the point of view of the freely falling observer, the field is in vacuum,
\begin{equation} n_\omega = \sum_{\omega '} \big| B_{\omega \omega '}' \big|^2 =
\frac{1}{e^{-2 \pi \kappa_-^{-1}\omega} - 1}.
\end{equation}
From this distribution one may infer that the temperature concerning the
particle radiation emitted
by the inner horizon is, when the redshift effects are ignored,
\be \label{eq:rupu}T = -\frac{\kappa_-}{2\pi},
\ee
which is \emph{positive}.

The result in Eq. (\ref{eq:rupu}) is in agreement with the
findings of Ref. \cite{pad}. Since the temperature is positive, there
are no interpretative problems concerning the thermodynamical properties of
the radiation of the inner horizon.
Again, the redshift factor can be recovered by using
Eq. (\ref{eq:tolman}), and as the result one finds that the
temperature, from the point of view of an observer at rest very close
to the inner horizon, is
\begin{equation} \label{eq:osku}T_- :=
-\frac{\kappa_-}{2\pi \Delta} = \frac{\sqrt{M^2-Q^2}}{2\pi r_-^2
  \sqrt{1-\frac{2M}{r} +\frac{Q^2}{r^2}}}.
\end{equation}

As we can see, our expression for the temperature of the particles
emitted by the inner horizon inside the inner horizon is very similar
to Eq. (\ref{eq:dilli2}), which gives the temperature of the particles
emitted by the outer horizon outside the outer horizon. The only
difference is that $r_+$ has been replaced by $r_-$. Our result that
the inner horizon emits particles inside the inner horizon with a
positive temperature given by Eq. (\ref{eq:osku})
has a most important consequence: To maintain a
local energy balance it is necessary that when real particles with
energy $\omega$ are emitted towards the singularity from the inner horizon,
\emph{antiparticles} with energy $-\omega$ are emitted \emph{away from
  the singularity} through the
inner horizon. The process is similar to the one which, according to
the Hawking effect, takes place at the outer horizon of the \RN black
hole: At the outer horizon antiparticles go in and particles come out,
and now we found that this is true
at the inner horizon as well. According to our best knowledge
this phenomenon, despite of its apparent triviality, has not been
noticed before.

\begin{figure}
\centering
\input{whitehole.pstex_t}
\caption{World lines of the particle and the antiparticle created at
  the inner horizon. The real particle meets with the black hole
  singularity, whereas the antiparticle travels accross the intermediate
  region between the horizons.}
\label{fig:luuppi}
\end{figure}
An intriguing question now arises: What happens to the antiparticles
which are emitted away from the singularity through the inner horizon? A hint for the answer to this
question may be found if we consider the conformal diagram of the
maximally extended \RN spacetime of Fig. \ref{fig:luuppi}. In that
figure we have drawn the world lines of the particle and the
antiparticle which are created at the inner horizon. The real particle,
as we found in our analysis, remains inside the inner horizon, and
finally meets with the singularity of the \RN hole. The antiparticle,
in turn, enters the intermediate region between the horizons and one may---if
the backscattering effects are neglected---speculate on the
possibility that it travels accross the
intermediate region and finally comes out from the
\emph{white hole horizon}. The situation is, however, quite
complicated since the vacuum states corresponding to a freely
falling observer near the inner horizon of \RN black hole and the white
hole horizon are completely different. Therefore, a detailed analysis
of the antiparticle modes travelling through the intermediate region
is needed.

Let us consider antiparticle solutions with energy $-\omega$ moving
towards the inner horizon in the region VI' of the
Fig. \ref{fig:luuppi}. We shall adhere a convention where the
Kruskal type coordinates defined at the neighborhoods of the points
$P_2$ and $P_1'$ are labelled by minus and plus sign, respectively. In that
case the antiparticle solutions may be written as
\be \label{eq:sisaan5} \phi ' \approx N_{\omega lm} Y_{lm}\, \frac{1}{r}\,  e^{i\omega \widetilde{u}_-},
\ee
where we have defined
\be \widetilde{u}_- := l_- (v_- - u_-).
\ee
The \RN metric is an analytic function of the coordinates $u_-$ and $v_-$ at the inner horizon, and
therefore the antiparticle solutions have exactly the same form as in
Eq. (\ref{eq:sisaan5}), when they are transferred accross the horizon
to the region IV'. The definitions of $u_-$ and $v_-$, however, change
according to the Eqs. (\ref{ozzy}), and in the region IV' we
have:
\begin{subequations}
\begin{eqnarray}
\label{eq:hesa} 
\left\{ \begin{array}{l}
u_- = \frac{1}{2} \big( e^{\kappa_- V} - e^{-\kappa_- U} \big), \\ 
v_- =\frac{1}{2} \big( e^{\kappa_- V} + e^{-\kappa_- U} \big),  \end{array}
\right.  \\
\label{eq:hesa2} 
\left\{ \begin{array}{l}
u_+ = -\frac{1}{2} \big( e^{\kappa_+ V} - e^{-\kappa_+ U} \big), \\ 
v_+ = -\frac{1}{2} \big( e^{\kappa_+ V} + e^{-\kappa_+ U} \big).  \end{array}
\right.
\end{eqnarray}
\end{subequations}
In general, the relationships between the coordinates (\ref{eq:hesa})
and (\ref{eq:hesa2}) are quite complicated. Therefore, for the sake of
simplicity, we confine ourselves to a special case where the \RN
black hole is almost extreme. More precisely, we shall assume that
\be \epsilon := \frac{r_+ -r_-}{r_+} < < 1.
\ee
In that case we have
\be \kappa_- (\epsilon ) = \frac{-\epsilon}{2r_+ (1 -\epsilon )^2},
\ee
and $\kappa_-$ can be expanded as series with respect to $\epsilon$
resulting
\be \kappa_- (\epsilon ) = - \kappa_+ + \mathcal{O} (\epsilon^2).
\ee
By using the above result one easily finds that
\be u_- = \frac{-u_+}{u_+^2 - v_+^2} + \mathcal{O} (\epsilon^2),
\ee
\be v_- = \frac{v_+}{u_+^2 - v_+^2} + \mathcal{O} (\epsilon^2),
\ee
and, moreover, 
\be \widetilde{u}_- = \frac{l_-}{u_+ - v_+} + \mathcal{O} (\epsilon^2).
\ee
When we define
\be \widetilde{u}_+ := l_+ (v_+ - u_+)
\ee
we finally get
\be \widetilde{u}_- = -\frac{l_- l_+}{\widetilde{u}_+ } + \mathcal{O} (\epsilon^2).
\ee
Therefore, at the region IV', one may write Eq. (\ref{eq:sisaan5}) as:
\be \label{eq:sisaan6} \phi ' \approx N_{\omega lm}
Y_{lm}\, \frac{1}{r}\,  e^{-i\omega \frac{l_- l_+}{\widetilde{u}_+ }},
\ee
when $\epsilon$ is sufficiently small.

When the solutions in Eq. (\ref{eq:sisaan6}) are transferred from the
region IV' into the
region I', they have again exactly the same form as above but the
coordinates $u_+$ and $v_+$ are defined differently according to
Eqs. (\ref{ozzy}). The antiparticle solutions coming out from the white
hole near the horizon in the region I' are, according to the observer
at rest with respect to the horizon,
\be \label{eq:sisaan7} \phi \approx N_{\omega lm} Y_{lm}\, \frac{1}{r}\,  e^{i\omega U}.
\ee
From Eq. (\ref{eq:makkara}) one finds that
\be
U = - \kappa_+^{-1} \ln (-\widetilde{u}_+) + \kappa_+^{-1} \ln l_+.
\ee
Thus we arrive at the Bogolubov transformation
\begin{equation}  e^{-i\omega \kappa_+^{-1}\ln (-\widetilde{u}_+)}\, e^{i\omega \kappa_+^{-1} \ln l_+} = \sum_{\omega '} \Big( A_{-\omega \omega '}'
e^{-i\omega'\frac{l_- l_+}{\widetilde{u}_+}} + B_{-\omega \omega '}' e^{i\omega '\frac{l_- l_+}{\widetilde{u}_+}} \Big).
\end{equation}
When one denotes
\be z := \frac{l_- l_+}{\widetilde{u}_+}
\ee
the above transformation becomes to:
\begin{equation}  e^{i\omega \kappa_+^{-1}\ln (-z)}\, e^{-i\omega \kappa_+^{-1} \ln l_-} = \sum_{\omega '} \Big( A_{-\omega \omega '}'
e^{-i\omega' z} + B_{-\omega \omega '}' e^{i\omega ' z} \Big).
\end{equation}
The Bogolubov coefficients can be obtained from the integrals
\begin{subequations}
\begin{equation} A_{-\omega \omega '}' = \frac{1}{2\pi} e^{-i\omega
    \kappa_+^{-1} \ln l_-} \int\limits_{-\infty}^0 dz\,  
e^{i\omega \kappa_+^{-1}\ln (-z)} e^{i\omega 'z},
\end{equation}
\begin{equation} B_{-\omega \omega '}' = \frac{1}{2\pi} e^{-i\omega \kappa_+^{-1} \ln l_-}\int\limits_{-\infty}^0 dz\, 
e^{i\omega \kappa_+^{-1}\ln (-z)} e^{-i\omega 'z}.
\end{equation}
\end{subequations}
These integrals lead to the result
\begin{equation} \big| A_{-\omega \omega '}' \big| = e^{\pi \kappa_+^{-1}\omega} \big| B_{-\omega \omega '}' \big|.
\end{equation} 
Therefore, by using Eq. (\ref{eq:pernaa}), we see that the
number of antiparticles with energy $-\omega$ is, from the point of view of
the observer at rest very close to the white hole horizon,
when, from the point of view of the freely falling observer, the field is in vacuum,
\begin{equation} \label{eq:finni} n_{-\omega} = \sum_{\omega '} \big| B_{-\omega \omega '}' \big|^2 =
\frac{1}{e^{2 \pi \kappa_+^{-1}\omega} - 1}.
\end{equation}
This is the Planck spectrum at the temperature
\be T = \frac{\kappa_+}{2\pi}
\ee
corresponding to the antiparticle flow coming out from the white hole. 

When redshift effects are taken into account, one finds that the
temperature of the white hole horizon for antiparticles is 
\begin{equation} T_{\text{WH}} :=
\frac{\sqrt{M^2-Q^2}}{2\pi r_+^2
  \sqrt{1-\frac{2M}{r} +\frac{Q^2}{r^2}}}.
\end{equation}
In other words, our analysis predicts that
not only do the black hole horizons of the \RN spacetime emit thermal
radiation with a black body spectrum but thermal radiation is emitted
by the white hole horizons as well. This means that outside the \RN
black hole there
exists two simultaneous radiation processes. They are the normal
black hole radiation, and the ``white hole radiation'' which is caused
by the pair creation effects at the inner horizon. The ``white hole
radiation'' predicted by our analysis is a new effect which according
to our best knowledge has not been mentioned in the literature
before. However, the white hole radiation does not consist of
particles with positive energy, but of antiparticles with negative energy.
The emission of antiparticles out of the white hole, in turn, may be
understood as an absorption of energy by the white hole horizon. This feature
contradicts with the classical results (classically, no energy may be
absorbed by the white hole horizon) in a similar way as does the
evaporation process at black hole horizons.

In our analysis we considered the special case where the \RN black
hole was almost extreme. This was indeed a very strong assumption but
it was needed for analytical calculations. It
is, however, natural to expect that there exists a certain antiparticle flow from a
white hole horizon even when the hole is quite far from
extremality. In that case one may also expect that the energy spectrum of the
antiparticles differs from the spectrum (\ref{eq:finni}), thus leading to
a different temperature as well. These kinds of effects may take place
especially when the absolute value of the electric charge $Q$ is, in
natural units, much smaller than $M$. 

\section{Concluding Remarks} \label{sec:con}
In this paper we have found that in maximally extended \RN spacetime
both the black and the white hole horizons emit thermal radiation
which, when the possible backscattering effects are neglected, obeys
the normal black body spectrum. The analysis implying an existence of the white
hole radiation was based, for the sake of simplicity, on the
assumption that the \RN black hole was almost extreme. However, it is
natural to expect that this phenomena also exists for black holes
far from extremality. In that case, the
situation becomes much more complicated and one may anticipate corrections to the
temperature of the white hole horizon. The radiation coming out from the
white hole horizons is caused by the pair creation effects at the
inner horizon of the \RN black hole. When a particle-antiparticle pair
is created just inside the inner horizon of the \RN hole, the
real particle is emitted towards the singularity from the inner horizon, whereas the
antiparticle is emitted away from the singularity through the inner horizon. The
particle finally meets with the black hole singularity, whereas the
antiparticle travels accross the intermediate region between the
horizons of the \RN black hole, and finally comes out from the white
hole horizon. In other words, we have found that outside the \RN black hole
there occurs two simultaneous radiation processes which are caused by
the pair creation effects at the both horizons of the hole. Since the
white hole radiation consists of antiparticles, the white hole radiation
process may be understood as an absorption of energy by the white hole horizon.

We obtained our results by means of an analysis which was similar to
the normal derivation of the Unruh effect. More precisely, we
considered the quantum-mechanical properties of the massless
Klein-Gordon field in the vicinity of the horizons from the points of
views of two different observers. One of these observers was at rest
with respect to the \RN coordinates either just outside the outer
horizon or inside the inner horizon, whereas another observer was in a
free fall through the horizon. We found that an observer at rest
observes particles even though, from the point of view of an observer
in a free fall, the field is in vacuum. The observer at rest just
outside the outer horizon observes outcoming real particles. Similarly
an observer at rest just inside the inner horizon observes
particles propagating towards the singularity.

The most remarkable result of our analysis is that, in contrast to
common beliefs, the inner horizon is not a passive spectator but an
active participant in the radiation processes of the \RN black
hole \cite{mr}. Although this result is based on an almost trivial observation
that both of the horizons of the \RN black hole emit particles,
there may also be some element of
surprise in it, and therefore the first question concerns the physical
and the mathematical validity of our analysis. After all, we did not
follow the usual route with an analysis based on a comparison of the
solutions to the Klein-Gordon equation in the past and in the future
light-like infinities. Since this kind of an analysis would have been
impossible when considering the radiation emitted by the inner
horizon, we instead compared the solutions in the rest frames of two
observers. Is this kind of approach valid?

The physical validity of this kind of approach has been considered,
in the case of the Schwarzschild black hole, by Unruh, and similar
arguments also apply here \cite{unruh}. The best argument in favor of the validity
of our analysis is probably given by the fact that exactly the same
methods which were used in an analysis of the radiation of the inner
horizon produced the well-known results for the radiation emitted by
the outer horizon. Another problem is that we have simply ignored all
possible backscattering effects. To consider such effects one should
perform a numerical analysis of the solutions to the Klein-Gordon
equation. However, as mentioned in Introduction, it is expected that the backscattered particles
will trigger the mass inflation, and therefore the radiation effects
of the inner horizon of the \RN black hole are of very brief
duration. As a consequence, the full \RN spacetime, even as a
mathematical solution, is unstable at the semiclassical limit. Since
the \RN spacetime is not considered astrophysically relevant, we do not
expect that our analysis has direct astrophysical consequences. If, however, our
results are qualitatively the same for more realistic Kerr
black holes, the phenomena discussed in this paper might possess some
astrophysical implications. Nevertheless, the radiation effects of the inner horizon
have much importance in their own right since they support the idea
that all horizons of spacetime emit radiation.

\begin{acknowledgements}
We thank Jorma Louko and Markku Lehto for their constructive criticism
during the preparation of this paper. This work was supported in part
by the Finnish Cultural Foundation.
\end{acknowledgements}

\appendix*
\section{Infinitesimal Geodesic Coordinates Near the Horizons
  of the \RN Spacetime}
Consider geodesic coordinates near the outer horizon of
a \RN black hole. More precisely, consider a geodesic
coordinate system in the region I infinitesimally close to the point $P_1$ of
Fig. \ref{fig:lasse}. By using the definitions
(\ref{roisto}) such that $\alpha = \kappa_+$ in
Eqs. (\ref{ozzy}), the metric
in the region I of \RN spacetime can be written as 
\be ds^2 = \label{eq:rekku2} \frac{1}{l_+^2}\frac{1}{\kappa_+^2 r^2} e^{-2\kappa_+ r} (r-r_-
)^{\frac{r_-^2}{r_+^2}+1} \big(-\big( dX^0\big)^2 +\big( dX^1\big)^2\big) + r^2(d\theta^2 + \sin^2
\theta\, d\varphi^2).
\ee 
Especially, on the two-surface $r=r_+$ the metric takes a very simple form: 
\be \label{eq:pallo} ds^2 =-\big( dX^0 \big)^2 +\big( dX^1 \big)^2 + r^2(d\theta^2 + \sin^2 \theta\, d\varphi^2),
\ee
Thus, the coordinates $X^0$ and $X^1$ provide an infinitesimal geodesic
coordinate system for the freely falling observer if only the derivatives of the metric with respect
to $X^0$ and $X^1$ vanish at the point $P_1$. According to Eq. (\ref{eq:rekku2}) the components of
the metric depend merely on $r$. Moreover, in order to show that the
derivatives of the metric vanish, it is sufficient to show that 
\be \label{eq:riisi} \frac{\partial r}{\partial u} = \frac{\partial r}{\partial v} = 0
\ee
at the point $P_1$.

Let us first note that the relationship between the
coordinates $u$, $v$, and $r$ can be expressed in an implicit form such that
\be \label{eq:previi} u^2 - v^2 = e^{2\kappa_+ r_*}.
\ee
By differentiating both sides with respect to $u$, one gets
\be \label{eq:viikate} 2u = 2\kappa_+ \, e^{2\kappa_+ r_*} \frac{dr_*}{dr} \frac{\partial
  r}{\partial u},
\ee
and similarly differentiation with respect to $v$ gives
\be \label{eq:viikate2} -2v = 2\kappa_+ \, e^{2\kappa_+ r_*} \frac{dr_*}{dr} \frac{\partial
  r}{\partial v}.
\ee
Since all equations (\ref{ozzy})
must be satisfied at $P_1$, one finds that $u = v = 0$ at $P_1$.
However, it is easy to see that when $r=r_+$,
\be 2\kappa_+ \, e^{2\kappa_+ r_* } \frac{dr_*}{dr} = 2 \kappa_+ r_+^2
\, e^{2\kappa_+ r_+} (r_+ - r_-)^{-\big( \frac{r_-^2}{r_+^2} +1 \big)}
\neq 0.
\ee
Hence Eq. (\ref{eq:riisi}) is satisfied at the point $P_1$.

Next, let us concentrate on the geodesic coordinates near the inner
horizon. Consider a geodesic
coordinate system in the region VI', infinitesimally close to the point $P_2$ of
Fig. \ref{fig:lasse2}. In this
case, we choose $\alpha = \kappa_-$ for the Kruskal-type coordinates.
By using the definitions (\ref{verneri}) and (\ref{eq:worf2}) the metric in the region VI' can be written as
\be ds^2 = \frac{1}{l_-^2} \frac{1}{\kappa_-^2 r^2} e^{-2\kappa_- r} (r_+ -r
)^{\frac{r_+^2}{r_-^2}+1} \big( -\big( dX^0 \big)^2 +\big( dX^1 \big)^2) + r^2(d\theta^2 + \sin^2
\theta\, d\varphi^2).
\ee 
When $r=r_-$, the metric has the form of that of
Eq. (\ref{eq:pallo}), and, therefore, the coordinates $X^0$ and $X^1$ 
provide an infinitesimal geodesic coordinate system if the derivatives
of the metric vanish at the point $P_2$. Again, one easily sees that it is sufficient to
show that Eq. (\ref{eq:riisi}) holds also at $P_2$. 

To show this, we proceed as before. The relationship between the
coordinates $u$, $v$, and $r$ can now be expressed as
\be u^2 - v^2 = e^{2\kappa_- r_*}.
\ee
Differentiating both sides with respect to $u$ and $v$ gives, respectively,
\be 2u = 2\kappa_- \, e^{2\kappa_- r_*} \frac{dr_*}{dr} \frac{\partial
  r}{\partial u},
\ee
\be -2v = 2\kappa_- \, e^{2\kappa_- r_*} \frac{dr_*}{dr} \frac{\partial
  r}{\partial v}.
\ee
Similarly as before, we have $u = v = 0$ at $P_2$. Furthermore, one
easily finds that, when $r=r_-$,
\be 2\kappa_- \, e^{2\kappa_- r_* } \frac{dr_*}{dr} = 2 \kappa_- r_-^2
\, e^{2\kappa_- r_-} (r_+ - r_-)^{-\big( \frac{r_+^2}{r_-^2} +1 \big)}
\neq 0.
\ee
Hence Eq. (\ref{eq:riisi}) is satisfied also at the point $P_2$.

\end{document}